\documentclass[a4paper,11pt]{article}
\usepackage{pos}

\title{Out of the Dark: WISPs in String Theory and the Early Universe}

\author*[a]{Joseph P. Conlon}

\affiliation[a]{Rudolf Peierls Centre for Theoretical Physics,\\
 Beecroft Building, Parks Road, OX1 3PU, Oxford, UK}

\emailAdd{joseph.conlon@physics.ox.ac.uk}

\abstract{New light hidden sector degrees of freedom represent one of the most approaches to going beyond the Standard Model. I give a short account of how such WISP candidates naturally appear in string compactifications and some descriptions of ways that they can affect early universe cosmology.}

\FullConference{1st General Meeting and 1st Training School of the COST Action COSMIC WISPers (COSMICWISPers)\\ 
5-14 September, 2023\\
Bari and Lecce, Italy\\}

\begin{document}
\maketitle

\section{Introduction}

The Standard Models of particle physics and cosmology are one of the triumphs of human understanding. Nonetheless, one of the central aims in science is to go beyond these and
understand the new physics, new interactions and new particles that supplement these. In discovering these, two complementary frontiers exist.

The first is the traditional high-energy frontier. The history of the Standard Model is replete with particles for which the principle obstacle to their discovery was simply that of energy. Such particles interact via either the electromagnetic or strong interaction; once they are made, they are relatively easy to observe. The difficulty lies in making them in the first place: their large masses places stringent kinematic obstacles in the way of their production, which can only be overcome through large accelerator complexes. In recent years, the top quark and the Higgs boson have been the paradigmatic example of particles discovered in this fashion.

The second frontier is the weak coupling frontier. On this frontier, the barrier to discovery is not that of energy but rather that of coupling strength. The difficulty here is not energy -- particles along this frontier may be extremely light and not energetically hard to make. Instead, here the obstacle is that when particles interact only extremely weakly, it is hard to detect either the presence or absence of such particles. Historically, the chief example of such a particle was the neutrino, which only interacts via the weak interactions. Although not a quantised particle as such, gravitational waves also represent an example of such phenomena, where all obstacles to production are energetic in nature.

WISPs belong firmly on the second frontier. This article describes how WISPs can naturally arise in string compactifications, how they may be generated in cosmology, and possible strategies to detect them.

\section{WISPs in String Theory}

String theory is the leading and most developed account of a theory that could unify gravitational and non-gravitational forces into a single framework. Recent reviews of string theory as applied to cosmology and particle physics are \cite{Cicoli:2023opf} and \cite{Marchesano:2024gul} respectively. If string theory is true, the world is ultimately described by a string compactification and the particle physics and spectrum present in four dimensions arises from the geometry, topology and field backgrounds present in the extra-dimensional geometry of the compactification. It is a generic feature of such geometries that the particle spectrum they give rise to involves WISP candidates; in this respect, one can say that the existence of WISPs is well motivated from string compactifications.

Here we detail several particular WISP candidates and the ways they arise in string compactifications.
\subsection{Axions and axion-like particles (ALPs)}

The potential for the strong CP problem of QCD to be solved through the existence of an axion has been recognised for a long time (a good broad overview of axion physics is given by \cite{Reece:2023czb}). Viewed from the perspective of 4-dimensional quantum field theory, the defining characteristic of an axion is its periodicity in field space. In contrast to the real scalar field $\phi(x, t)$ encountered in a typical \emph{Canonical Quantum Field Theory 101 for Poets} course, an axion field $a(x, t)$ is valued on a circle with a finite field range,
\begin{equation}
a(x, t) \equiv a(x, t) + 2 \pi f_a,
\end{equation}
where $f_a$ is the axion decay constant. 

This makes it clear that a non-trivial topology (in field space) is essential to the idea of an axion: as the vev of the field changes, it eventually returns to its original value. Such topological complexity is also a key feature of string compactifications, where all the physics that we would see in four dimensions arises from the structure of the extra dimensions. The Calabi-Yau spaces typically present in string compactification have much topological structure, counted by the Hodge numbers $h^{1,1}$ and $h^{2,1}$. These represent the number of non-trivial 2-cycles and 3-cycles in the compactification; they count the K\"ahler and complex structure moduli respectively (see \cite{Greene:1996cy} for a review of Calabi-Yau manifolds).

These non-trivial topological structures give rise to axions in the effective field theory. The simplest case is that of world-sheet axions. The string action 
\begin{equation}
S \sim \int \sqrt{g} e^{-2 \phi} + B_2 + \dots
\end{equation}
involves a coupling of the 2-form $B_2$ to the worldsheet. The 2-form decomposes on dimensional reduction as $B_2 = b_i \omega^i$, where $\omega^i$ is the 2-form associated to the different 2-cycles in the compactification (numbered by $i$). In the path integral, $e^{iS}$ is insensitive to $2 \pi$ variations and so $b_i$ correspond to axions in the low-energy field theory.

A similar behaviour is encountered in type IIB and type IIA theories for Ramond-Ramond axions arising from dimensional reduction of the $C_2$, $C_3$ and $C_4$ form field on cycles within the Calabi-Yau. The RR fields only enter the low-energy action via the D-brane action, 
\begin{equation}
S  = \int \sqrt{g} e^{-\phi} + i \int_{\Sigma_j} C_j
\end{equation}
where $\Sigma_j$ is a cycle and $C_j$ the corresponding RR form, and again result in periodic field (axions) in the low-energy effective field theory.

\subsection{Hidden photons and other dark gauge groups}

A second type of WISP that naturally arises in string theory are hidden photons (a review of dark photons is \cite{Fabbrichesi:2020wbt}). Unless they are Higgsed, the massless nature of such vector bosons is protected by the gauge symmetry. As a massless vector has two polarisation states and a massive vector three polarisation states, quantum corrections are not sufficient by themselves to give a mass to an otherwise massless gauge boson. Dark gauge sectors with minimal couplings to the Standard Model are relatively unconstrained. Direct bounds tend to arise via kinetic mixing of such a dark U(1) boson with the photon of electromagnetism (observational limits on the kinetic mixing parameter are reviewed in  \cite{Fabbrichesi:2020wbt}). Furthermore, although dark sector U(1)s are most commonly considered, in principle non-Abelian dark gauge groups are also possible, with the mass of excitations related to the confinement scale in the hidden sector (which can be exponentially light, in analogous to the behaviour of the strong force in the Standard Model).

String compactifications can generate such dark sector gauge groups in at least two distinct ways, both related to the geometry of the compactification. The first comes from the presence of D-branes (either individual or stacks) in the compact space. As is well know, a stack of $N$ D-branes supports an $SU(N)$ gauge group, with a single D-brane supporting a $U(1)$ gauge group. In the case that a D-brane is present at an orbifold or orientifold singularity, then the brane can \emph{fractionate} and be viewed as a composite object made up of various fractional branes each supporting different gauge groups. In this way, it is possible for branes (or stacks of branes) to support a variety of complex gauge and matter sectors.

In many string models, the Standard Model can be realised as a \emph{local construction} : all the degrees of freedom contributing to the Standard Model are localised at a particular point within the compactification geometry. The physics of this is particularly clear in weakly coupled D-brane models, as the mass of an extended stringy mode grows with its length. If the Standard Model sector is physically separated from a distant brane stack, then any modes charged under both visible and hidden gauge sectors have string-scale masses: such a sector is decoupled from the visible sector and so represents a WISP sector.

Another way hidden sector U(1)s can arise in string theory is via dimensional reduction of higher-dimensional form fields such as the RR $C_3$ and $C_4$ fields. When a 3-form or 4-form is reduced on a 3-cycle or 4-cycle (respectively), they give rise to axions. When reduced on 2-cycles or 3-cycles (respectively), they give rise to 1-form fields: i.e. U(1) vector bosons. So, in the presence of topological structure (ubiquitous in Calabi-Yaus), hidden gauge bosons arise as e.g
\begin{equation}
A^i_{\mu} = \int_{\Sigma_3^i} C_4,
\end{equation}
where $\Sigma_3$ is a 3-cycle (assumed to survive any orientifold projection).

\subsection{Hidden sector chiral fermions}

Although their phenomenological importance is not widely studied, another class of fields which can give rise to light, weakly coupled particles are hidden sector chiral fermions. In this case, chirality protects the mass: a chiral fermion cannot acquire a mass without some form of Higgs mechanism. In phenomenology, such particles can arise in gauge sectors that are decoupled from the Standard Model. The neutrino is a visible sector chiral fermion with (effectively) vanishing mass. Hidden sectors analogues of the neutrino would be phenomenological examples of hidden sector chiral fermions.

In string theory, hidden sector chiral fermions can arise in a analogous way to hidden sector U(1)s. In the case of two separated D3 branes at singularities, there will be two separate gauge groups, each with chiral matter, present in the compactification. Chiral fermions present on one gauge group are hidden from the perspective of the other, and so will correspond to massless dark sector fermions.

\section{WISPs in Early Universe Cosmology}

There are therefore various ways WISPs can naturally appear in string compactifications. How can these have an impact on observational physics? We focus on cosmology and, in particular, string-motivated cosmologies.

\subsection{WISPs from moduli decay: A Cosmic WISP Background}

One of the most prominent ways in which string theory tends to modify early universe cosmology is through a long period of moduli domination. This arises as the stringy moduli are non-relativistic matter particles which only interact via gravitational strength interactions. The former property means that they come to dominate over ambient radiation, as their energy density redshifts as $\rho_m \sim a^{-3}$ in contrast to the $\rho_{\gamma} \sim a^{-4}$ of radiation. The latter property ensures that they live a long time as all their interactions arte suppressed by a factor of $M_P$. In particular, dimensional grounds imply that the characteristic lifetime of a modulus $\Phi$ is
\begin{equation}
\tau_{\Phi} \sim 8 \pi \frac{M_P^2}{m_{\Phi}^3} \sim 10^{-6} {\rm s} \left( \frac{10^6 {\rm GeV}}{m_{\Phi}} \right)^3
\end{equation}
Given inflation can end at $10^{-35} {\rm s}$ the moduli-dominated epoch can last an exceedingly long time.

In string compactifications, reheating occurs at the time of moduli decay. As moduli are gravitationally coupled (and in the case of e.g. the volume modulus may be sensitive to all fields
in the compactification), when they decay their decay products can include all of the fields present in the compactification. Furthermore, there is no reason for Standard Model fields to be parametrically preferred. This may result in a significant branching ratio of the modulus into light WISP degrees of freedom, such as axions or dark photons. Such WISPs are formed with an initial energy of $E = m_{\Phi}/2$ which then redshifts through the expansion of the universe, giving rise to a Cosmic WISP background (studied for the case of axions in \cite{Cicoli:2012aq, Higaki:2012ar}). This represents a stringy approach to producing an isotropic and homogeneous background of relativistic WISPs in cosmology.

\subsection{WISP enhancement via kination epochs}

Kination epochs in the early universe are ones where the universe is dominated by the kinetic energy of a scalar field. They can naturally arise in string theory from epochs where scalar fields evolve towards runaway directions in field space (for example, towards the large volume or weak coupling limits). In a kination epoch, the universe traverses approximately one Planckian distance in field space each Hubble time: making them particularly well suited to being studied in a stringy context, as transPlanckian field excursions cannot be controlled within low-energy effective field theory alone: it requires knowledge of the UV completion to control Planck-suppressed operators.

During a kination epoch, the overall energy density of the universe redshifts as
\begin{equation}
\rho_{kin} \sim a(t)^{-6}.
\end{equation}
This implies that, during a kination epoch, even small initial amounts of radiation present in the universe will catch up with the dominant kinating component. In the context of a scalar field on an exponential potential, the endpoint will often be a tracker solution with the energy shared between kinetic energy, potential energy and background radiation.

This implies that even small amounts of initial background WISP radiation can be enhanced during a kination epoch so that it becomes an $\mathcal{O}(1)$ fraction of the energy density of the universe: this amplification can make such WISP radiation backgrounds much more prominent than they would otherwise be, generating epochs of the universe where they form an $\mathcal{O}(1)$ fraction of the energy density of the universe. A detailed recent study of such epochs in stringy cosmology is \cite{Apers:2024ffe}.

\section{Conclusions}

String theory is the most promising candidate for a fundamental theory of nature combining gauge theories and gravity in a single consistent quantum framework. String compactifications naturally contain many well-motivated WISP candidates: axions, dark photons and dark fermions. String compactifications also motivate various modifications to the standard cosmology in which WISP candidates can play a prominent role and may potentially give observable signatures.

\section*{Acknowledgements}

This article is based upon a talk delivered at the 1st General Meeting of the COSMIC WISPers COST Action CA21106, supported by COST (European Cooperation in Science and Technology). I also acknowledge support from the STFC consolidated grants ST/T000864/1 and ST/X000761/1

\end{document}